\documentclass[twocolumn,prl,superscriptaddress]{revtex4}

\usepackage{dcolumn}
\usepackage{amsmath}
\usepackage{calrsfs}

\usepackage{graphicx}

\newcommand{\half}{\tfrac12}
\newcommand{\etal}{{\it{}et~al.}}
\newcommand{\defn}{\textit}
\newcommand{\Ord}{\mathrm{O}}
\renewcommand{\Im}{\mathop\mathrm{Im}}
\newcommand{\mat}{\mathbf}
\renewcommand{\vec}{\mathbf}
\newcommand{\from}{\leftarrow}

\begin{document}

\title{Equitable random graphs}
\author{M. E. J. Newman}
\affiliation{Department of Physics and Center for the Study of Complex
  Systems,
University of Michigan, Ann Arbor, MI 48109}
\author{Travis Martin}
\affiliation{Department of Electrical Engineering and Computer Science,
  University of Michigan, Ann Arbor, MI 48109}

\begin{abstract}
  Random graph models have played a dominant role in the theoretical study
  of networked systems.  The Poisson random graph of Erd\H{o}s and R\'enyi,
  in particular, as well as the so-called configuration model, have served
  as the starting point for numerous calculations.  In this paper we
  describe another large class of random graph models, which we call
  equitable random graphs and which are flexible enough to represent
  networks with diverse degree distributions and many nontrivial types of
  structure, including community structure, bipartite structure, degree
  correlations, stratification, and others, yet are exactly solvable for a
  wide range of properties in the limit of large graph size, including
  percolation properties, complete spectral density, and the behavior of
  homogeneous dynamical systems, such as coupled oscillators or epidemic
  models.
\end{abstract}

\maketitle

In the rapidly growing branch of physics devoted to the study of networks,
random graphs are probably the most widely studied class of model systems.
They are the ``Ising model'' of networks, idealized systems that capture
the crucial features of real networks while remaining simple enough to be
solvable, exactly or approximately, for many properties of interest.
Random graphs have been employed in countless calculations on networks over
the years, from classic works in the 1950s~\cite{SR51,ER59} up to the
present day.

There are several different random graph models in wide current use.  The
simplest is the Poisson random graph of Erd\H{o}s and
R\'enyi~\cite{ER59,Bollobas01} in which edges are placed uniformly and
independently at random among a set of vertices.  A~more realistic example
is the configuration model, in which the degrees of vertices are fixed but
connections between them are in other respects random~\cite{MR95,NSW01}.

In this paper we describe a further large class of random graphs, which we
call \defn{equitable random graphs} and which are suitable as models for a
wide range of structures found in networked systems.  Equitable random
graphs are flexible enough to permit any choice of vertex degrees, while
allowing us to incorporate many other features into the network, such as
community structure or bipartite structure.  At the same time, as we will
demonstrate, equitable random graphs are exactly solvable for a broad range
of static and dynamic properties.  We will give example solutions of three
specific properties: spectral density of the adjacency matrix, percolation
properties, and the dynamics of homogeneous dynamical systems of coupled
equations.

While equitable random graphs are, like other random graph models, a
simplified representation of the structure found in real-world networks,
they have the potential to provide a flexible and powerful starting point
for the mathematical study of the interplay between the structure and
behavior of networked systems.

We define an equitable random graph as follows: $n$~vertices are divided
into $q$ groups and undirected edges are placed between them such that each
vertex in group~$\mu$ has $m_{\mu\nu}$ edges to vertices in group~$\nu$,
where $n$, $q$, and $m_{\mu\nu}$ are parameters whose values we choose.
Apart from the constraint on the numbers of edges between groups, edges are
placed at random.  Note that in general $m_{\mu\nu}$ is not symmetric:
$m_{\mu\nu} \ne m_{\nu\mu}$.

One can think of the equitable random graph as a variant on the widely
studied stochastic block model~\cite{HLL83,KN11a}, in which vertices are
divided into $q$ groups and edges placed between them with probabilities
$p_{\mu\nu}$ that depend on group membership.  The equitable random graph
is similar, but fixes the number of edges between groups rather than their
probability.  It is also similar in spirit to, though different in
important details from, the object known as the (non-stochastic) block
model, which has a long history of study in sociology~\cite{WF94}.

An alternative way of looking at the equitable random graph---and the one
that inspired our naming of it---is that it is a graph drawn at random from
the set of graphs with a given equitable partition.  An equitable partition
is precisely a division of a graph's vertices into some number of groups
such that all vertices in a group have the same numbers of connections to
each group.  Equitable partitions are commonly used, for example, in
computer algorithms for graph isomorphism.

Equitable random graphs are capable of representing many types of network
structure.  As a simple example, we could generate an equitable random
graph with two equally sized groups of $\half n$ vertices each and specify
that each vertex in group~1 has $a$ neighbors in group~1 and $b$ in
group~2, while each vertex in group~2 has $b$ neighbors in group~1 and $a$
in group~2.  If $a>b$ the result is a random network displaying ``community
structure''---groups of vertices with more connections within groups than
between them~\cite{GN02,Fortunato10}.  It is straightforward to generalize
this kind of structure to more than two groups.  Alternatively, we could
set $a<b$, placing more edges between groups than within them and creating
the inverted community structure known as disassortative mixing.  In the
extreme case $a=0$ in which vertices have no connections at all to their
own group we get a bipartite graph, another structure type that has been
widely studied in the literature.  As a further example, consider an
equitable random graph with a large number of groups labeled by $r =
1\ldots q$, and suppose vertices in each group have connections only to
their own group~$r$ and to the immediately adjacent groups~$r\pm1$.  This
produces what is called a ``stratified'' network in the social networks
literature---a network with layers in which each layer is connected only to
the adjacent ones.  For instance, friendship networks are sometimes
observed to be stratified by age, individuals primarily being friends with
those of the same age, or a little older or younger.

Given the values of the model parameters, constructing an equitable random
graph, for instance on a computer, is a straightforward process.  The only
mildly challenging part is working out the number~$n_\mu$ of vertices in
each group~$\mu$, which is not stated explicitly in our definition of the
model.  We note that the number of edges from vertices in group~$\mu$ to
vertices in group~$\nu$ is $n_\mu m_{\mu\nu}$, while the number running in
the opposite direction is $n_\nu m_{\nu\mu}$.  But, the edges being
undirected, these two numbers are necessarily equal: $n_\mu m_{\mu\nu} =
n_\nu m_{\nu\mu}$ for all $\mu,\nu$.  Summing over~$\nu$, dividing by~$n$,
and defining a $q\times q$ mixing matrix~$\mat{M}$ with elements
$m_{\mu\nu}$ and a vector $\vec{x}$ with elements $x_\mu = n_\mu/n$, we
find that $\vec{x}^T = \vec{x}^T\mat{M}\mat{D}^{-1}$, where $\mat{D}$ is
the diagonal matrix with elements $D_{\mu\mu} = \sum_\nu m_{\mu\nu}$.

In other words, the vector~$\vec{x}$, whose elements represent the fraction
of vertices in each of the groups, is a left eigenvector of
$\mat{M}\mat{D}^{-1}$, and moreover it must be the eigenvector associated
with the largest (most positive) eigenvalue, by the Perron--Frobenius
theorem, since all elements of $\mat{M}\mat{D}^{-1}$ are non-negative.
Once we have~$\vec{x}$, the number of vertices in each group~$\mu$ is
simply~$nx_\mu$.  In practice these numbers need not be exact integers and
we may have to round to the nearest integer, but this rounding will
introduce a vanishing error in the limit of large network size, which is
the primary regime of interest for random graphs.

Once the number of vertices is fixed, one can create the random graph itself
by giving each vertex an appropriate number of ``stubs'' of edges, each
labeled with the group to which it is meant to attach, then picking
compatible stubs in pairs at random and joining them to make complete
edges.  The end result is a matching of the stubs drawn uniformly at random
from the set of all possible matchings.

Our main interest in equitable random graphs, however, is not in numerical
studies and computer simulation, but in their analytic study.  In the
remainder of this paper we give exact solutions for several properties of
the networks generated by the model.  For our first example, we study the
percolation properties of equitable random graphs.

Percolation is the process of activating or ``occupying'' a fraction of the
vertices or edges of a network, chosen at random, then looking at the
structure of the subgraph consisting of the occupied entities.  Percolation
is widely used as a model for the robustness of networks to the failure of
vertices or edges~\cite{CEBH00,CNSW00} and as a model of the spread of
epidemics~\cite{Grassberger83,Newman02c}.  Here we consider the case of
edge (or bond) percolation, in which the edges of the network are occupied
independently at random with some probability~$p$ that we choose.  In
general, one finds that for small~$p$ the occupied edges form only small
clusters of connected vertices but when $p$ passes a certain threshold
value~$p_c$, called the percolation threshold, the clusters coalesce to
form a giant or percolating cluster that fills a nonzero fraction of the
network (with the remainder of the network still being divided into small
clusters).

A crucial property of equitable random graphs for our purposes is that they
are ``locally tree-like.''  Since edges are placed randomly, the
probability of their forming a loop of any finite length in the network
vanishes in the limit of large~$n$, and the neighborhood of any node looks
like a tree.  As shown in~\cite{KNZ14}, the percolation properties of
locally tree-like networks can be expressed in terms of a message passing
process.  Vertex~$i$ receives a message~$u_{i\from j}$ from its
neighbor~$j$ equal to the probability that $i$ is \emph{not} connected to
the percolating cluster via vertex~$j$, and the messages satisfy the
self-consistent condition
\begin{equation}
u_{i\from j} = 1 - p + p \prod_{k\in\mathcal{N}(j)\backslash i}
               u_{j\from k},
\label{eq:uij}
\end{equation}
where the notation $\mathcal{N}(j)\backslash i$ denotes the set of
neighbors of vertex~$j$ excluding vertex~$i$.  Then the probability~$v_i$
that vertex~$i$ itself is not in the percolating cluster is
\begin{equation}
v_i = \prod_{j\in\mathcal{N}(i)} u_{i\from j},
\label{eq:vi}
\end{equation}
and the expected size~$S$ of the percolating cluster as a fraction of~$n$
is given in terms of the average of these probabilities over all vertices
by $S = 1 - (1/n) \sum_i v_i$.

For an equitable random graph the crucial observation is that every edge
from group~$\mu$ to group~$\nu$ has the same neighborhood---the pattern of
edges around it is exactly the same as for every other edge from $\mu$
to~$\nu$, out to arbitrarily large distance on a large graph.  This means
that there exists a solution for the messages such that~$u_{i\from j}$
depends only on the groups to which~$i$ and $j$ belong and not on the
individual vertex labels.  Then Eqs.~\eqref{eq:uij} and~\eqref{eq:vi}
become
\begin{equation}
u_{\mu\from\nu} = 1 - p + p \prod_\lambda
  u_{\nu\from\lambda}^{m_{\nu\lambda}-\delta_{\lambda\mu}},\quad
v_\mu = \prod_\nu u_{\mu\from\nu}^{m_{\mu\nu}},
\label{eq:umunu}
\end{equation}
where $\delta_{\lambda\mu}$ is the Kronecker delta.  On a network with $m$
edges in total, this reduces the original set of $2m$ messages~$u_{i\from
  j}$ to a much smaller set of size~$q^2$.

These equations are already useful as a tool for numerical
computation---they can be solved by simple iteration, starting from a
random initial condition, far more quickly than the full equation
set~\eqref{eq:uij}.  But we can also use them for exact analytic
calculations of percolation properties.  Consider, for example, the
percolation threshold~$p_c$.  Following an argument of~\cite{KNZ14}, we
note that as $p$ approaches~$p_c$ from above, all
messages~$u_{\mu\from\nu}$ approach~1 (since the
probability~$1-u_{\mu\from\nu}$ of $\mu$ being connected to the percolating
cluster vanishes below~$p_c$ because there \emph{is} no percolating
cluster).  Just above the threshold, therefore, $u_{\mu\from\nu} = 1 -
\epsilon_{\mu\from\nu}$ for some small~$\epsilon_{\mu\from\nu}$ and,
expanding~\eqref{eq:umunu} to leading order, we find that
\begin{equation}
\epsilon_{\mu\from\nu} = p \sum_\lambda
  \bigl( {m_{\nu\lambda}-\delta_{\lambda\mu}} \bigr)
  \epsilon_{\nu\from\lambda}.
\label{eq:percolation}
\end{equation}
We can rewrite this in matrix notation as $\boldsymbol{\epsilon} =
p\mat{B}\boldsymbol{\epsilon}$, where $\boldsymbol{\epsilon}$ is the
$q^2$-element vector with elements~$\epsilon_{\mu\from\nu}$ and $\mat{B}$
is a $q^2\times q^2$ real non-symmetric matrix with elements indexed by
$\mu\from\nu$ and equal to
\begin{equation}
B_{\mu\from\nu,\kappa\from\lambda} = \delta_{\kappa\nu}
   \bigl( m_{\kappa\lambda} - \delta_{\lambda\mu} \bigr).
\end{equation}
Equation~\eqref{eq:percolation} then tells us that the percolation
threshold~$p_c$ is equal to the reciprocal of the leading eigenvalue
of~$\mat{B}$---and it must be the leading eigenvalue, by the
Perron--Frobenius theorem once again, since $\epsilon_{\mu\from\nu}$ has
all elements non-negative.

Consider as an example the two-group network discussed earlier in which
every vertex has $a$ connections to its own group and $b$ connections to
the other group.  In this case, the matrices $\mat{M}$ and~$\mat{B}$ take
the form
\begin{equation}
\mat{M} = \begin{pmatrix}
            a & b \\ b & a
          \end{pmatrix}, \qquad\setlength{\arraycolsep}{5pt}
\mat{B} = \begin{pmatrix}
            a-1 & b   & 0   & 0   \\
            0   & 0   & b-1 & a \\
            a   & b-1 & 0   & 0   \\
            0   & 0   & b   & a-1
          \end{pmatrix},
\end{equation}
and the leading eigenvalue of~$\mat{B}$ is $a+b-1$.  Hence
\begin{equation}
p_c = {1\over a+b-1}.
\label{eq:pc}
\end{equation}
Indeed for the general case of $q$ identically sized groups with each
vertex having $a$ in-group connections and $b$ connections to every other
group it is not hard to show that the percolation threshold falls at $p_c =
1/[a + (q-1) b - 1]$.  One can also solve in a straightforward manner for
the size of the percolating cluster and the average size of the small
clusters.

\begin{figure}
\begin{center}
\includegraphics[width=\columnwidth]{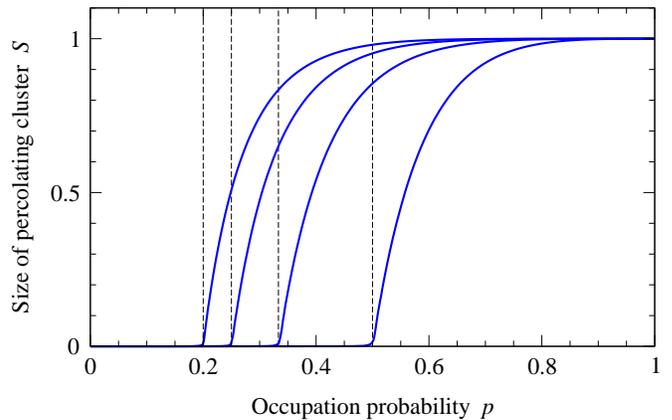}
\end{center}
\caption{Size of the percolating cluster in numerical simulations of bond
  percolation on two-group networks of the type described in the text with
  $n=1\,000\,000$ nodes each.  From left to right, the four curves have
  $(a,b) = (4,2)$, $(3,2)$, $(3,1)$, and~$(2,1)$.  The vertical dashed
  lines represent the theoretical predictions for the position of the
  percolation threshold, Eq.~\eqref{eq:pc}.}
\label{fig:perc}
\end{figure}

Figure~\ref{fig:perc} shows measurements of the size of the percolating
cluster on computer-generated equitable random graphs with a range of
different choices for the parameters $a$ and~$b$.  As the figure shows, the
position of the percolation transition agrees well in each case with our
analytic predictions (indicated by the vertical dashed lines).

For our second example of solvable properties of equitable random graphs,
we consider the spectral density of the adjacency matrix.  Adjacency matrix
spectra find uses in defining centrality measures, in algorithms for graph
partitioning and community detection, in network visualization, and
numerous other areas.  The spectral density~$\rho(z) = (1/n) \sum_i
\delta(z-\lambda_i)$ is the probability density of eigenvalues~$\lambda_i$
and, as recently shown by Rogers~\etal~\cite{RCKT08}, it can be calculated
on locally tree-like graphs using a message passing technique.  In the
formulation we use (which differs slightly from that of Rogers~\etal)\ the
messages are functions~$g_{i\from j}(z)$ satisfying
\begin{equation}
g_{i\from j}(z) = {1\over1 - z\sum_{k\in\mathcal{N}(j)\backslash i}\>
                   g_{j\from k}(z)},
\label{eq:gij}
\end{equation}
in terms of which the spectral density is
\begin{equation}
\rho(z) = - {z\over n\pi}
\Im \sum_i {1\over z^2 - \sum_{j\in\mathcal{N}(i)}\> g_{i\from j}(1/z^2)}.
\label{eq:rho}
\end{equation}

Again the crucial observation for equitable random graphs is that all edges
between vertices in groups~$\mu$ and~$\nu$ have the same neighborhood out
to arbitrary distances and hence that there exists a solution such that the
message~$g_{i\from j}(z)$ depends only on the groups to which $i$ and~$j$
belong and not on the specific vertices.  Equation~\eqref{eq:gij} then
simplifies to
\begin{equation}
g_{\mu\from\nu}(z) = {1\over1 - z\sum_\lambda
        (m_{\nu\lambda} - \delta_{\lambda\mu}) g_{\nu\from\lambda}(z)},
\label{eq:gmunu}
\end{equation}
which reduces the problem of calculating the spectral density from one of
solving $\Ord(m)$ equations to one of solving a fixed number~$q^2$ even in
an arbitrarily large system.

\begin{figure}
\begin{center}
\includegraphics[width=\columnwidth]{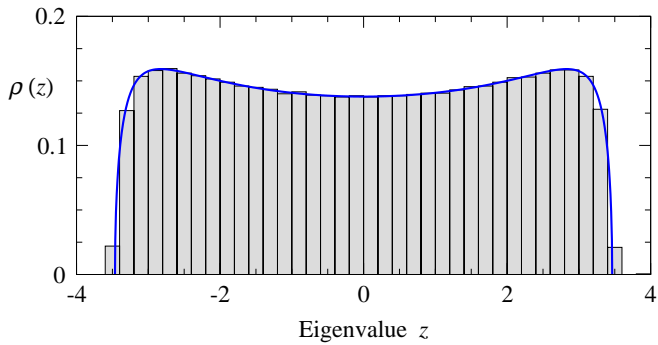}
\end{center}
\caption{Spectrum of the two-group equitable random graph described in the
  text.  The solid curve represents the analytic solution,
  Eq.~\eqref{eq:km}, for the case $a=3$, $b=1$, while the histogram shows
  the results from the direct numerical diagonalization of a single
  computer-generated instance with $n=10\,000$ nodes and the same parameter
  values.}
\label{fig:spectrum}
\end{figure}

Taking once again the example of the two-group model with community
structure discussed previously, there are in principle four different
functions~$g_{\mu\from\nu}(z)$, but because of symmetry between the groups
all four are equal in this case, and Eq.~\eqref{eq:gmunu} reduces to a
single equation:
\begin{equation}
g(z) = {1\over1-(a+b-1)zg(z)}.
\end{equation}
Solving the resulting quadratic and substituting the solution
into~\eqref{eq:rho}, we find the spectral density to be
\begin{equation}
\rho(z) = {(a+b)\over2\pi}\, {\sqrt{4(a+b-1)-z^2}\over (a+b)^2-z^2},
\label{eq:km}
\end{equation}
which is a version of the Kesten--McKay distribution.
Figure~\ref{fig:spectrum} shows an example spectrum for the case $a=3$,
$b=1$ along with numerical results from direct diagonalization of the
adjacency matrix for a computer-generated equitable random graph with the
same parameters, and the agreement is excellent.  Note that although the
graph is extremely sparse in this case, the calculation still gives a
correct result, by contrast with some other calculations of graph spectra,
which fail in the sparse limit~\cite{RB88,Kuhn08}.

For our third example calculation with equitable random graphs we consider
the evolution of a dynamical system on a network.  In particular, we
consider homogeneous systems, in which degrees of freedom on each vertex
obey the same dynamics, and vertices are coupled along the edges of the
network.  Widely studied examples include coupled oscillators on networks
and the dynamics of epidemic disease models such as the
susceptible--infected--recovered (SIR) model.

Let us denote by $x_i, y_i, \ldots$ the degrees of freedom of a
dynamical system on vertex~$i$ of a network, let $\vec{r}_i =
(x_i,y_i,\ldots)$, and let us write the dynamics in the standard form
\begin{equation}
\dot{\vec{r}}_i = \vec{f}(\vec{r}_i) +
                  \sum_j A_{ij} \vec{g}(\vec{r}_i,\vec{r}_j),
\label{eq:dynamics}
\end{equation}
where $A_{ij}$ is an element of the adjacency matrix.  Here
$\vec{f}(\vec{r})$ represents the intrinsic dynamics of the vertex (which
is the same for every vertex), and $\vec{g}(\vec{r}_i,\vec{r}_j)$
represents the effect of vertex~$j$ on vertex~$i$.  (Note that $\vec{g}$ is
not necessarily symmetric in its arguments.)  Systems with dynamics
governed by second- or higher-order differential equations do not fit this
form, but one can always reduce a set of second-order equations to twice as
many first-order equations by introducing auxiliary variables, and
similarly for higher orders.

As discussed by, for example, Golubitsky and Stewart~\cite{GS06}, the size
or complexity of dynamical systems on networks can in some cases be reduced
significantly by exploiting network symmetries.  Equitable random graphs do
not typically possess significant symmetries, but nonetheless similar
reductions are possible.  We focus on the case where all vertices have the
same initial condition (or, more generally, all vertices in each group of
the random graph have the same initial condition).  In this case, all
vertices in each group will evolve in an identical fashion, because each,
by definition, has the same intrinsic dynamics and feels the same influence
from its neighbors.  Thus our $n$ equations~\eqref{eq:dynamics} immediately
reduce to just~$q$:
\begin{equation}
\dot{\vec{r}}_\mu = \vec{f}(\vec{r}_\mu)
                    + \sum_\nu m_{\mu\nu} \vec{g}(\vec{r}_\mu,\vec{r}_\nu).
\label{eq:sir}
\end{equation}

As an example, consider the SIR model of epidemic disease, whose state can
be represented by $\vec{r}_i = (s_i,x_i,r_i)$ on each vertex with
$s_i,x_i,r_i$ denoting the probability that the vertex is, respectively,
susceptible to, infected with, or recovered from the disease of interest at
a given time, and $\vec{f}(\vec{r}_i) = (0,-\gamma x_i,\gamma x_i)$,
$\vec{g}(\vec{r}_i,\vec{r}_j) = (-\beta s_i x_j, \beta s_i x_j, 0)$, where
$\beta$ and~$\gamma$ are constants representing the rate of infection and
recovery per unit time respectively.  For the equitable random graph we
then have
\begin{align}
\dot{s}_\mu &= -\beta s_\mu \sum_n m_{\mu\nu} x_\nu, \\
\dot{x}_\mu &= \beta s_\mu \sum_n m_{\mu\nu} x_\nu - \gamma x_\mu, \\
\dot{r}_\mu &= \gamma x_\mu,
\end{align}
in the approximation where all vertex states are considered independent.
We can apply any of the standard techniques for nonlinear systems to this
set of equations---finding fixed points, linear stability analysis,
bifurcation analysis, or in some cases exact solutions.  As an example, we
can in this case make progress by eliminating $x_\mu$ between the first and
third equations and integrating to get $s_\mu =
\exp\bigl[-(\beta/\gamma)\sum_\nu m_{\mu\nu} r_\nu\bigr]$.  At long times
the disease always dies out, so that $x_\mu=0$, and combining this with the
fact that $s_\mu+x_\mu+r_\mu=1$ we have
\begin{equation}
r_\mu = 1 - \exp\biggl(-{\beta\over\gamma} \sum_\nu m_{\mu\nu} r_\nu\biggr)
\end{equation}
as $t\to\infty$.  The solution to this equation tells us the number of
recovered individuals at long times in each group, which is necessarily
equal to the number of individuals who ever had the disease---in other
words it tells us the size of the disease outbreak, group by group.  We
note that the equation always has a trivial solution at $r_\mu=0$ for
all~$\mu$, which corresponds to a situation in which there is no epidemic.
It may or may not have a nontrivial solution, corresponding to the presence
of an epidemic, depending on whether the rate of infection~$\beta$ is large
enough compared to the rate of recovery~$\gamma$.  Expanding the equation
for small values of~$r_\mu$ (i.e.,~close to the regime where there is no
epidemic) and rearranging, we get $\sum_\nu m_{\mu\nu} r_\nu =
(\gamma/\beta) r_\mu$, or $\mat{M}\vec{r} = (\gamma/\beta) \vec{r}$ in
matrix notation.  In other words, at the transition point $\vec{r}$~is the
leading right eigenvector of the mixing matrix~$\mat{M}$ and $\gamma/\beta$
is the corresponding eigenvalue.  To put that another way, there will be an
epidemic if, and only if, $\gamma/\beta$ is less than the leading
eigenvalue of the mixing matrix.  (A result reminiscent of this one is seen
in epidemic behavior in stochastic block models---see~\cite{BP02a}.)

As an example, consider a network with two equally sized groups and mixing
matrix
\begin{equation}
\mat{M} = \begin{pmatrix}
            4a & 2a \\
            2a & a \\
          \end{pmatrix}.
\end{equation}
This model divides the network into a dense core (group~1) and a sparse
periphery (group~2) with connections between them of intermediate density.
Core--periphery structure of this kind is widely observed in real-world
networks~\cite{BE99,Holme05b,RPFM14}.  Since the leading eigenvalue of this
mixing matrix is~$5a$, we can immediately see that the epidemic threshold
falls at $\gamma/\beta = 5a$.

In summary, we have in this paper described a broad class of random graph
models, which we call equitable random graphs, that is flexible enough to
represent complex structure types such as community structure,
core--periphery structure, and stratification, which have historically been
of interest in the study of networked systems.  At the same time these
models can be solved for a range of nontrivial structural and dynamic
properties, including percolation properties, graph spectra, and behavior
of homogeneous dynamical systems on their vertices.  This combination of
flexibility and solvability gives equitable random graphs the potential to
substantially enhance our understanding of the interplay between structure
and function in networks.

The authors thank Brian Ball and Martin Golubitsky for useful
conversations.

\end{document}